\documentclass[pra,twocolumn]{revtex4}
\usepackage[english]{babel}
\usepackage{amsmath}
\usepackage{amssymb}
\usepackage{epsfig}
\usepackage{url}

\begin{document}
\title{Nonlinear tunnel oscillations of light in spherical Bragg resonators}
\author{Victor P. Ruban}
\email{ruban@itp.ac.ru}
\affiliation{Landau Institute for Theoretical Physics RAS,
Chernogolovka, Moscow region, 142432 Russia}

\date{\today}

\begin{abstract}
A weakly nonlinear regime of radial tunneling is theoretically considered
for a light wave in a spherical dielectric Bragg resonator containing a set
of ``shell'' eigenmodes (both TE and TM) with different azimuthal numbers
$l\geqslant 1$, which are concentrated near a defect of the Bragg structure,
several layers away from the origin. Also, a single mixed radial mode with
$l=1$ is located near the resonator center, either TM or TE. Nonlinearity of
Kerr type in the main approximation is actual for the central mode only,
while all the shell modes remain in the linear regime. The tunneling occurs
between the central mode and the shell $l=1$ mode of the same symmetry.
Nontrivial part of the dynamics of optical field is described by a Hamiltonian
system of ordinary differential equations for complex vectors ${\bf C}_{1}(t)$
and ${\bf C}_{2}(t)$, which determine the magnitude and spatial orientation
of the wave structures at the center and at the shell, respectively.
Depending on ``asymmetry parameter'' of tunnel coupling, the system
demonstrates different variants of nonlinear behavior.

\vspace{1mm}


\vspace{1mm}

\end{abstract}

\maketitle

\section{Introduction}

Wave tunneling is a fundamental phenomenon observed in many fields of physics.
As known, when a weak coupling between two resonators with nearly equal
eigen-frequencies takes place, tunnel oscillations are possible in such wave system.
For instance, in quantum mechanics in the simplest case the wave is of scalar type,
the motion is one-dimensional, and the resonators are potential wells
\cite{Theor_Phys_3}. For optics, however, the case of vectorial wave in
three-dimensional space is more important. Such problems are technically more
complicated (see, e. g., \cite{PRE65_056618,PRA76_051802_R,PhysRep2008,
PRL102_153901,L2009,PhysRep2012,PRL114_245503}). In particular, the light can
be localized on defects of some photonic crystals (if the forbidden
band exists there) and tunnel between such modes (see, e. g.,
\cite{J1987,HCS1990,YGMRBJ1991,Y1993,JVF1997Nature,BTO2000}, and references therein).
In this connection, of special interest are dielectric layered structures with
spherical symmetry, where the radial confinement of light is realized through
the Bragg reflection, and complete linear analysis is easy to perform due to
separation of variables (see \cite{BPS1993,BDdAZL2012,GPK2022,GPLK2023}, and
references therein). By a properly designed profile of dielectric permittivity
$\varepsilon(r,\omega)$, it is possible to realize close frequencies for trapped
states concentrated near defects of the $r$-periodic structure. In this case,
the eigen-frequencies fall into the forbidden band of radial motion.

In the linear approximation, such dielectric spherical Bragg resonators have been
well studied, especially for the cases where the permittivity profile
$\varepsilon(r,\omega)$ is a piecewise constant function of radial coordinate $r$
at fixed frequency $\omega$ \cite{BDdAZL2012,GPK2022,GPLK2023}. However, nonlinear
effects definitely require more attention. The purpose of this work is to take into
account nonlinearity of Kerr type in the dynamics of tunnel oscillations of light,
when two weakly coupled radial modes corresponding to the azimuthal number $l=1$
are present, one of them being located near the resonator center, while the other
mode occupying a ``shell'' several (about 10--20) layers away. A system of ordinary
differential equations (ODE) will be derived here for simplified description of
tunneling processes. Vectorial nature of light will be seen to fully manifest itself
in such system, particularly in the structure of nonlinear interaction. Due to a
quite small relative difference between the two eigen-frequencies, even a low
intensity of the optical field will be able to introduce strong nonlinear effects.
A ``parameter of asymmetry'' will appear in the ODE system, describing the
relationship between the true eigen-modes and their more localized linear
combinations. That parameter will make an additional contribution to a rich
variety of nonlinear dynamical regimes. Besides fundamental scientific interest,
the results obtained here can be useful for optical technologies.

\section{Structure of eigen-modes}

Let us first remind some necessary facts from the linear theory. We consider a
locally isotropic, spatially nonuniform dielectric medium (with temporal dispersion,
but without spatial dispersion) in its ``transparency window'' \cite{Theor_Phys_8}.
In other words, the imaginary part of dielectric permittivity in the frequency range
of our interest is negligibly small (of course, it is not small outside the
transparency window, i. e. in absorption bands). For practical applicability
of results obtained in this work, condition $\mbox{Im}\,\varepsilon\lesssim 10^{-4}$
is required, as it will be clear from the further consideration. It is not restrictive,
since the absorption of light can be actually much weaker, for example in glass.

Thus, in a sufficiently wide frequency interval near some $\omega_0$, we adopt
the model of a perfect dielectric with no material losses. Since the Kramers-Kronig
relations should be necessary satisfied, $\varepsilon(r,\omega)$ is a nontrivial
real positive function at real $\omega$ in the indicated domain, with inequality
$\partial\varepsilon/\partial\omega > 0$ taking place \cite{Theor_Phys_8}.

With the above assumptions, in the linear approximation the problem is reduced
to finding spatially localized solutions of the vector equation for the complex
amplitude of (monochromatic) electric field,
\begin{equation}
\mbox{curl}\,\mbox{curl}\,{\bf E}=\frac{\omega^2}{c^2}\varepsilon(r,\omega) {\bf E}.
\label{E_eq_lin}
\end{equation}
Two types of eigen-modes exist (see, e. g., Ref.\cite{BPS1993}).
Transverse-electric (TE) modes have the form
\begin{equation}
{\bf E}_{\rm TE}=\Psi(r)[\nabla_{\bf n} Y_{l,m}({\bf n})\times {\bf n}],
\end{equation}
where $Y_{l,m}$ is the spherical harmonics, ${\bf n}={\bf r}/r$ is the unit vector,
and a radial function $\Psi(r)$ satisfies the equation
\begin{equation}
-\Psi''-\frac{2}{r}\Psi'+\frac{l(l+1)}{r^2}\Psi
-\frac{\omega^2}{c^2}\varepsilon(r,\omega)\Psi=0.
\label{TE_eq}
\end{equation}
For transverse-magnetic (TM) modes, the following equality takes place,
\begin{equation}
\frac{c^2}{\omega^2}\mbox{curl}\,{\bf E}_{\rm TM}=
F(r)[\nabla_{\bf n}Y_{l,m}({\bf n})\times {\bf n}],
\end{equation}
so that the electric induction $\varepsilon(r,\omega) {\bf E}_{\rm TM}$
is given by formula
\begin{equation}
\varepsilon {\bf E}_{\rm TM} = l(l+1)\frac{F(r)}{r} {\bf n}Y_{l,m}({\bf n})
+\frac{d}{r dr}[r F(r)] \nabla_{\bf n}Y_{l,m}({\bf n}).
\end{equation}
The equation for function $F(r)$ is somewhat different from Eq.(\ref{TE_eq}):
\begin{equation}
-F''-\frac{2}{r}F'+\frac{l(l+1)}{r^2}F
+\frac{\varepsilon'}{\varepsilon r}\frac{d}{dr}(rF)
-\frac{\omega^2}{c^2}\varepsilon F=0,
\label{TM_eq}
\end{equation}
where $\varepsilon'=\partial\varepsilon/\partial r$.

In our case, the permittivity $\varepsilon(r,\omega)$ behaves periodically on $r$
at large distances from the origin. Therefore differential equations for functions
$\psi(r)=r\Psi(r)$ and $f(r)=rF(r)$ contain in that region the only non-periodic
coefficient $l(l+1)/r^2$, and it is negligibly small with moderate values of $l$.
For such equations, a gap in the frequency spectrum (the forbidden band) is known
to exist. When $\omega$ is inside the gap (the Bragg regime), the general solution
is a sum of two oscillatory functions with non-uniform amplitudes, one of them
exponentially growing and the other exponentially decaying.

We are interested in regular solutions $\Psi(r)$ and $F(r)$ which at small
$r\ll c/\omega_0$ behave as $r^l$, while at $r\to +\infty$ they tend to zero.
In the Bragg regime, amplitude of their oscillations approaches zero at large
$r$ exponentially fast (see examples in figures 1 and 2). Within transparency
window, such bound states exist for some special discrete (non-degenerate)
real values of frequency only, when the coefficient in front of the growing
exponent takes zero value. Since the eigen-modes $\Psi(r)$ and $F(r)$ are
determined up to an arbitrary multiplier, they can be taken purely real
without limitation of generality. Numerically, these functions can be found
by standard methods very easily. In particular, the ``shooting'' method works
quite accurately.

We should say that it was the neglect of material losses that allowed us to avoid
many difficulties related to the complexity of eigen-frequencies in the general case.
Another important point is that radiative losses can be considered as negligibly
small for a large but finite size $R_{\rm max}$ of the Bragg structure due to the
exponential decay of the modes at large $r$. Practically, the maximal radius as
several tens of wavelength is quite enough. If necessary, small material and
radiative losses can be included into analysis via a perturbative procedure.

\begin{figure}
\begin{center}
\epsfig{file=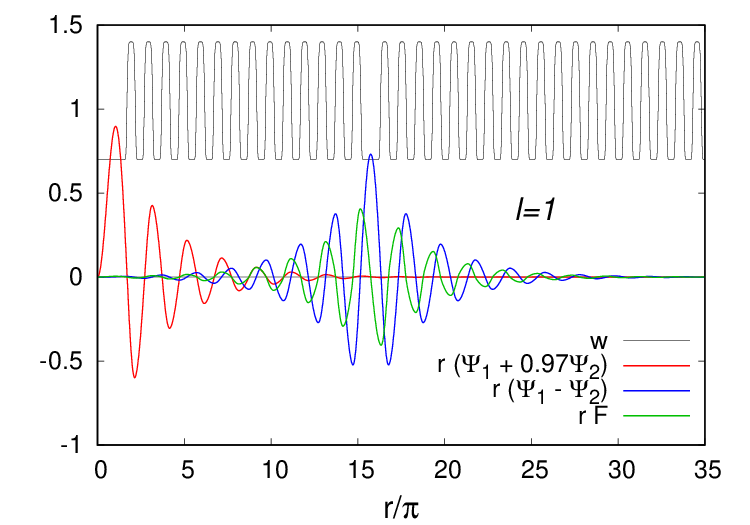, width=0.96\columnwidth}
\caption{
Examples of localized (non-normalized) linear solutions in the presence of central
TE-mode. Here 14 layers with larger dielectric permittivity are arranged through
distance $\pi$ starting from coordinate $r_0=1.938\pi$, and all the other layers
are shifted along $r$ additionally on value $0.612\pi$. The eigen-frequencies are
$\xi_1=-0.00066164155$, $\xi_2=0.001185237$. Also the shell mode of TM type is shown,
for which $\xi_{\rm TM}=0.0002669188$.
}
\label{TE}
\end{center}
\end{figure}

\section{Quasi-monochromatic approximation}

In this work, we will deal with modes having eigen-frequencies near $\omega_0$.
Therefore it will be convenient to normalize the time variable $t$ by factor
$1/\omega_0$, and the radial coordinate $r$ by factor $c/\omega_0\sqrt{\bar \varepsilon}$,
where $\bar \varepsilon$ is a value of dielectric permittivity somewhere in the middle
of the range between its minimum and maximum in the $r$-periodic profile at $\omega_0$.
We stress that any linear combination of such modes is quasi-monochromatic in the
frequency range, but not in the wave-vector space. Indeed, spatial Fourier-spectrum of
each mode is rather wide.

The crucial simplification in our further analysis will be achieved through expansion
of function $\omega^2\varepsilon(r,\omega)$ up to the first order on small deviation
$\xi=(\omega/\omega_0 -1)$, so that
\begin{equation}
\frac{\omega^2\varepsilon(r,\omega)}{\omega_0^2\bar\varepsilon}\approx w(r)+\tau(r)\xi.
\label{xi_appr}
\end{equation}
Since $\omega_0$ is assumed to lie in a transparency window of dielectric
medium, functions $w(r)$ and $\tau(r)$ are purely real and strictly positive.
Moreover, $\tau(r)>2 w(r)$ by virtue of the inequality
$\partial\varepsilon/\partial\omega > 0$ \cite{Theor_Phys_8}. So,
approximation (\ref{xi_appr}) reduces Eq.(\ref{E_eq_lin}) to a more-less standard
self-adjoint problem for eigen-values and eigen-modes. 

For numerical examples presented later, we used expression
\begin{eqnarray}
&&w(r)+\tau(r)\xi=0.7+1.5\xi\nonumber\\
&&\qquad\qquad+\sum_j(0.7+1.7\xi)\exp[-4.0(r-r_j)^6],
\label{eps_profile}
\end{eqnarray}
where the array of positive numbers $r_j$ determines radial coordinates of middles
of layers with higher dielectric permittivity. Such a smooth function to some extent
approximates a layered structure with sharp boundaries between the layers.
In the absence of ``defects'', successive values $r_j$ differ on $\pi$, thus forming
a spherical Bragg mirror, but existence of localized modes is possible only when at
least one interval is different from $\pi$. The widths of such ``anomalous'' layers
then determine the type of eigen-modes as well as their frequencies $\xi_\nu$. By a
careful choice of the intervals, all the values $\xi_\nu$ can be made quite small,
in accordance with the adopted quasi-monochromatic approximation
(see examples of profiles $w(r)$ with defects in figures 1 and 2).

\begin{figure}
\begin{center}
\epsfig{file=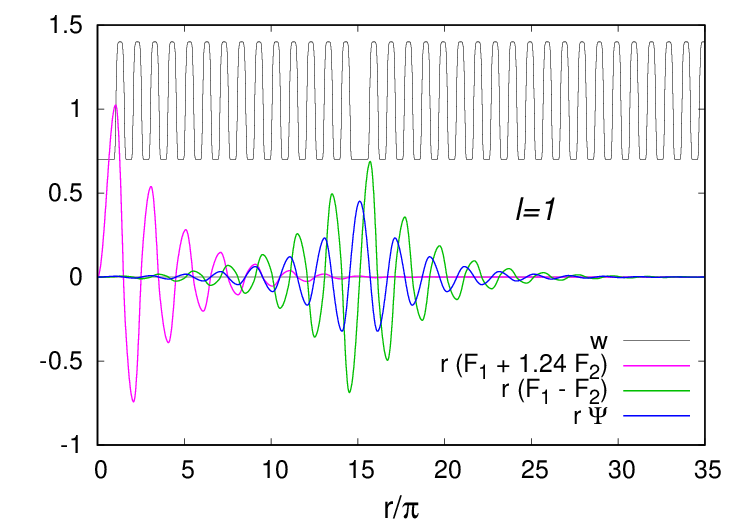, width=0.96\columnwidth}
\caption{
Examples of localized (non-normalized) linear solutions in the presence of central
TM-mode. Here 14 layers with larger dielectric permittivity are arranged through
distance $\pi$ starting from coordinate $r_0=1.302\pi$, and all the other layers
are shifted along $r$ additionally on value $0.62\pi$. Here the eigen-frequencies are
$\xi_1=-0.0023929839$, $\xi_2=-0.0001324124$. Also the shell mode of TE type is shown,
for which $\xi_{\rm TE}=-0.0013708785$.
}
\label{TM}
\end{center}
\end{figure}

In this work, the most simple case is considered when two radial modes with $l=1$
of the same symmetry (either $\Psi_1$ and $\Psi_2$, or $F_1$ and $F_2$) are present,
such that their localization is distributed between the central defect and a ``shell''
defect located at a distance $r_{\rm shell}$ about 10--20 layers (5--10 wavelengths)
away from the origin. All the other modes are completely concentrated at the ``shell''
defect. Actually, such a situation is not something special, since a central defect
size suitable to confine a TE mode is typically not good for trapping a TM mode, and
vice versa.

It is very important that there exist such linear combinations of eigen-modes
(for instance, $s_1=S_{11}\Psi_1+S_{12}\Psi_2$ and $s_2=S_{21}\Psi_1+S_{22}\Psi_2$),
that one of them is nearly completely concentrated around the center, and the other
on the ``shell''. The concept of such ``localization matrix'' $S_{\nu\mu}$ implies
optimization of some properly defined ''quality of localization''. In the simplest
case of two modes the choice of the combinations can be made roughly by eye.
Just such combinations are presented in figures 1 and 2. Recalling the famous
example with two potential wells from quantum mechanics, we clearly see here a
direct analogy between the ``one-well'' states and our ``mixed'' functions $s_1$,
$s_2$. With proper normalization, as it will be shown later, matrix $S_{\nu\mu}$
is orthogonal,
\begin{equation}
S_{\nu\mu}=\frac{1}{\sqrt{1+a^2}}\left(\begin{array}{cc}
1 &  a \\
a & -1 \end{array}\right),
\end{equation}
where $a$ is an ``asymmetry parameter'' depending on the dielectric permittivity
profile. Therefore coefficients $c_1(t)$ and $c_2(t)$ in expression for the radial
mode (with fixed number $m$),
\begin{equation}
\Psi_{\rm tot} e^{i\xi_1 t}=c_1 s_1+c_2 s_2=A_1(0)\Psi_1+A_2(0)\Psi_2 e^{-i\Delta t},
\end{equation}
where $\Delta=(\xi_2 -\xi_1)$, 
are governed by the following system of ordinary differential equations,
\begin{equation}
i\dot c_1=\frac{\Delta[a^2 c_1 -a c_2]}{1+a^2},\qquad
i\dot c_2=\frac{\Delta[ -a c_1  + c_2]}{1+a^2}.
\end{equation}

\section{Effect of nonlinearity}

It should be noted that local intensity of light for the ``central'' mixed mode
$s_1(r)$ is typically much larger than for $s_2(r)$ and all the other shell modes
(the corresponding large factor is about $(r_{\rm shell}/\pi)^2\sim$ 100--400).
Therefore, with moderate wave amplitudes, nonlinearity effectively acts in the
central part of the resonator only. The presence of just a single central radial
mode $s_1$ makes nonlinear analysis quite simple in comparison with the case of
simultaneous existence of several central radial modes. Indeed, in our situation
there will be no need to calculate terms of nonlinear cross-interactions between
different central states.

To take into account effects of Kerr nonlinearity, we should go back to the basic
equation
\begin{equation}
c^2\mbox{curl}\,\mbox{curl}\,{\bf E}({\bf r},t)=-\partial^2_t {\bf D}({\bf r},t)
\end{equation}
and include into $\partial^2_t {\bf D}$ the corresponding cubic terms in the
quasi-monochromatic approximation \cite{Theor_Phys_8}. As the result, we will
arrive at (dimensionless) equation
\begin{eqnarray}
\mbox{curl}\,\mbox{curl}\,{\bf E}&=&w(r) {\bf E}+i\tau(r)\partial_t {\bf E}\nonumber\\
&&+\alpha(r)|{\bf E}|^2{\bf E}+\beta(r)({\bf E}\cdot{\bf E}){\bf E}^*,
\label{E_eq_nonlin}
\end{eqnarray}
where $\alpha(r)$ and $\beta(r)$  are (real) Kerr coefficients, and ${\bf E}^*$ is
the complex conjugate vector.

It is important that Eq.(\ref{E_eq_nonlin}) represents a non-canonical
Hamiltonian system
\begin{equation}
i\tau(r)\partial_t {\bf E}=\delta{\cal H}/\delta{\bf E}^*,
\end{equation}
with the Hamiltonian functional ${\cal H}={\cal H}^{(2)}+{\cal H}^{(4)}$, where
\begin{eqnarray}
{\cal H}^{(2)}&=& \int\Big[|\mbox{curl}\,{\bf E}|^2-w(r)|{\bf E}|^2\Big]d{\bf r},\\
\label{H_2}
{\cal H}^{(4)}&=&-\int\Big[\frac{\alpha(r)}{2}|{\bf E}|^4
    +\frac{\beta(r)}{2}|({\bf E}\cdot{\bf E})|^2\Big]d{\bf r}.
\label{H_4}
\end{eqnarray}
In this case any two eigen-functions of linear approximation with differing frequencies
automatically turn out to be mutually orthogonal, if the scalar product is defined by
the following formula,
\begin{equation}
\langle 1 | 2\rangle=\int \tau(r) ({\bf E}_1^*\cdot{\bf E}_2) d{\bf r}.
\end{equation}
The angular dependences through $Y_{l,m}$ ensure the orthogonality also for $(2l+1)$-fold
degenerate states with definite $l$. At $l=1$, it will be convenient for us to use
real functions $({\bf e}_k\cdot {\bf n})$ as the angular multipliers, where $k=\{x,y,z\}$.

Tunnel processes are better to study in terms of the previously introduced functions
$s_1$ and $s_2$ (or analogous mixed modes in the case of TM states). In view of
orthonormality of the eigen-modes, the corresponding transformation matrix $S_{\nu\mu}$
can be taken orthogonal, which fact was mentioned earlier. In particular, for TE modes
the summed electric field is expressed as
\begin{equation}
{\bf E}_{\rm TE}e^{i\xi_1 t}= [{\bf C}_1(t)\times {\bf n}] s_1(r)
                             +[{\bf C}_2(t)\times {\bf n}] s_2(r),
\label{E_sum}
\end{equation}
and functions $s_\nu(r)$ are normalized as follows,
\begin{equation}
\frac{8\pi}{3}\int_0^\infty \tau(r) s_\nu^2(r) r^2 dr = 1.
\end{equation}

It is implied that forbidden band in the frequency spectrum is much wider than
$\Delta$ (with the permittivity profile corresponding to Eq.(\ref{eps_profile}),
we have $\Delta/\Delta_{\rm gap}\sim 0.01$). In such situation, nonlinear
interaction between trapped modes and continuous spectrum is much weaker than
between the confined states. Therefore, if the vectors ${\bf C}_{1,2}$ are both
of the same order and sufficiently small, then nonlinearity does not introduce
essential distortions into expression (\ref{E_sum}), but it just modifies
equation of motion for the coefficient ${\bf C}_1(t)$ comparatively to the
linear regime. In contrast, the shell-mode amplitude ${\bf C}_2(t)$ is not
affected directly by the nonlinearity in virtue of smallness (by a factor
$\sim \pi/r_{\rm shell}$) of the electric field there in comparison with the
central mode. The other shell modes with different $l$ remain in the linear
regime as well, and thus they do not participate in the dynamics.

Nonlinear corrections can be easily calculated by the variational method through
the Lagrangian structure of Eq.(\ref{E_eq_nonlin}) and orthogonality of modes
$s_1$ and $s_2$. It is sufficient to substitute (\ref{E_sum}) into expression
(\ref{H_4}), and thus obtain
\begin{equation}
H^{(4)}=-\frac{\tilde\alpha}{2}|{\bf C}_1|^4
        -\frac{\tilde\beta} {2}|({\bf C}_1\cdot{\bf C}_1)|^2,
\end{equation}
where the new effective Kerr coefficients $\tilde\alpha=(7 I_\alpha +2 I_\beta)$
and $\tilde\beta=(I_\alpha +6 I_\beta)$ are expressed through the integrals
\begin{equation}
I_\alpha=\frac{4\pi}{15}\int_0^{\infty}\alpha s_1^4 r^2 dr,
\quad
I_\beta =\frac{4\pi}{15}\int_0^{\infty} \beta s_1^4 r^2 dr.
\end{equation}
As the result of substitution of variational ansatz (\ref{E_sum}) into
the corresponding Lagrangian functional
${\cal L}=i\int \tau(r)({\bf E}^*\cdot\partial_t {\bf E})d{\bf r}-{\cal H}$,
the equations of motion for vectors ${\bf C}_1$ and ${\bf C}_2$ take the form
\begin{eqnarray}
i\dot {\bf C}_1&=&\frac{\Delta}{1+a^2}[a^2 {\bf C}_1 -a {\bf C}_2]\nonumber\\
&&-\tilde\alpha|{\bf C}_1|^2{\bf C}_1-\tilde\beta ({\bf C}_1\cdot{\bf C}_1) {\bf C}^*_1,
\label{C1_nonlin}
\\
i\dot {\bf C}_2&=&\frac{\Delta}{1+a^2}[-a {\bf C}_1  + {\bf C}_2].
\label{C2_nonlin}
\end{eqnarray}

An analogous system of ODE (but with different expressions for parameters
$\tilde\alpha$ and $\tilde\beta$) is obtained when the tunneling of TM modes
is considered. We do not present here the corresponding formulas. Equations
(\ref{C1_nonlin})-(\ref{C2_nonlin}) constitute the main result of this work.

It is essential that with small $\Delta\sim 0.001$, nonlinearity comes into
play already at quite moderate light intensities near the resonator center,
corresponding to relative nonlinear frequency shift as 
$\delta\omega_{\rm nl}/\omega_0\sim\alpha|{\bf E}|^2\sim\Delta$. Thus,
the physical limitation $\alpha|{\bf E}|^2\ll 1$ is definitely satisfied. 

Let us also note that passage to new variables $\tilde t=\Delta t$ and
${\vec C}_{1,2}=\sqrt{\tilde\alpha/\Delta} {\bf C}_{1,2}$ allows us to put formally
$\Delta=1$ and $\tilde\alpha =1$ in system (\ref{C1_nonlin})-(\ref{C2_nonlin}).
As the result, only two essential parameters remain: $a$ and
$\eta=\tilde\beta/\tilde\alpha$. In the numerical examples below $\eta=0.5$.

Integrals of motion for equations (\ref{C1_nonlin})-(\ref{C2_nonlin}), besides
the Hamiltonian
\begin{eqnarray}
H&=& \frac{1}{1+a^2}[a^2|{\vec C}_1|^2+|{\vec C}_2|^2
-a({\vec C}^*_1\cdot{\vec C}_2)-a({\vec C}^*_2\cdot{\vec C}_1)]\nonumber\\
&&-\frac{1}{2}|{\vec C}_1|^4 -\frac{\eta} {2}|({\vec C}_1\cdot{\vec C}_1)|^2,
\end{eqnarray}
are also the action (full intensity)
\begin{equation}
I=I_1+I_2=|{\vec C}_1|^2+|{\vec C}_2|^2,
\end{equation}
and the angular momentum
\begin{eqnarray}
{\vec M}&=&{\vec M}_1+{\vec M}_2\nonumber\\
&=&2[\mbox{Re}\,{\vec C}_1\times \mbox{Im}\,{\vec C}_1]
  +2[\mbox{Re}\,{\vec C}_2\times \mbox{Im}\,{\vec C}_2].
\end{eqnarray}

\begin{figure}
\begin{center}
\epsfig{file=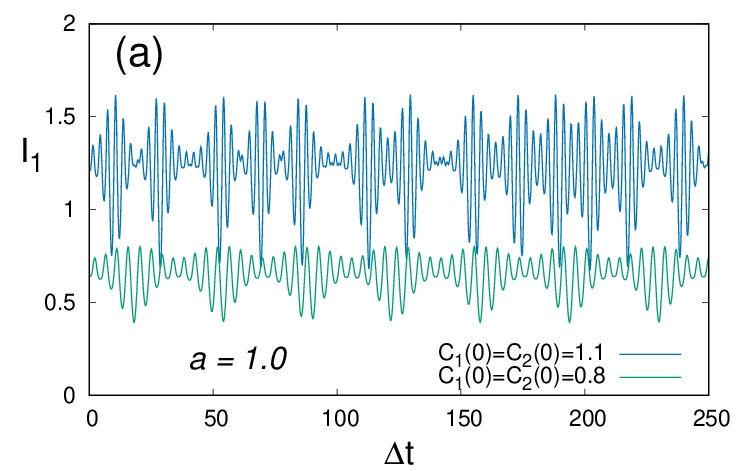, width=0.96\columnwidth}\\
\epsfig{file=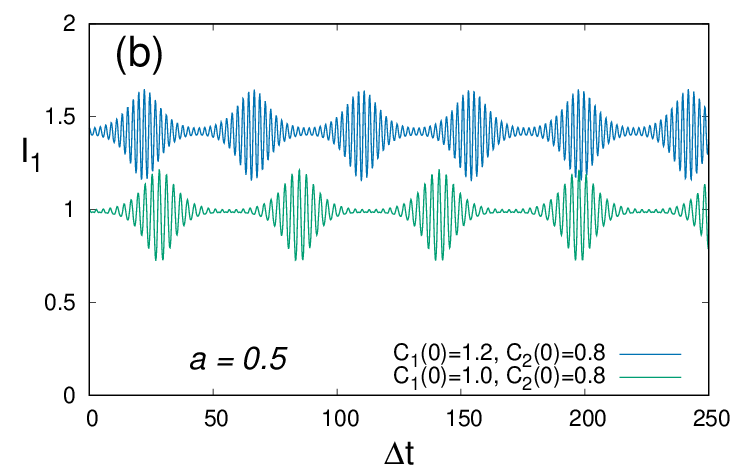, width=0.96\columnwidth}\\
\epsfig{file=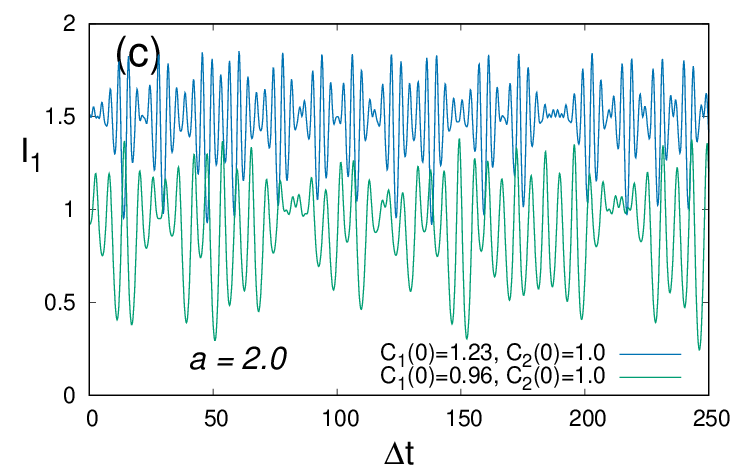, width=0.96\columnwidth}
\caption{
Examples of various nonlinear regimes in the modelling of radial tunneling
of light in the framework of equations (\ref{C1_nonlin})-(\ref{C2_nonlin}).
}
\label{C1C2_time}
\end{center}
\end{figure}

The presence of nonlinearity in system (\ref{C1_nonlin})-(\ref{C2_nonlin})
changes the character of radial tunneling dramatically in comparison with the
simple linear regime. In figure 3, numerical examples of temporal dependences of
intensity $I_1$, for several values $a$, are given with initial conditions
in the form ${\vec C}_1(0)=(C_1(0),0,0)^T$, ${\vec C}_2(0)=(0,C_2(0),0)^T$.
In some cases the dynamics reveals quasi-random features. Examination has
shown that exponential divergence of initially close phase trajectories
takes place there, which is a typical property of dynamical chaos.

In the final example presented in figure 4, with $a=2.0$, initial conditions
were ${\vec C}_1(0)=(0.47,0,0.2 i)^T$ and ${\vec C}_2(0)=(0,0.5,0.3 i)^T$.
The total angular momentum is not zero in this case, contrary to the previous
examples. We observe here a kind of activity burst accompanied by exchange
between the angular momenta.

\begin{figure}
\begin{center}
\epsfig{file=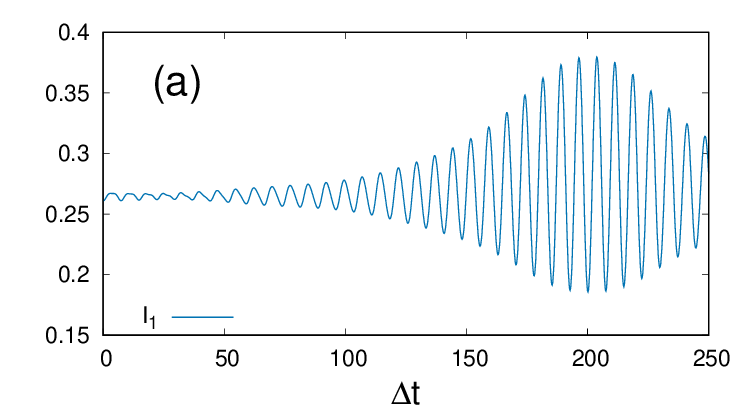, width=0.97\columnwidth}\\
\epsfig{file=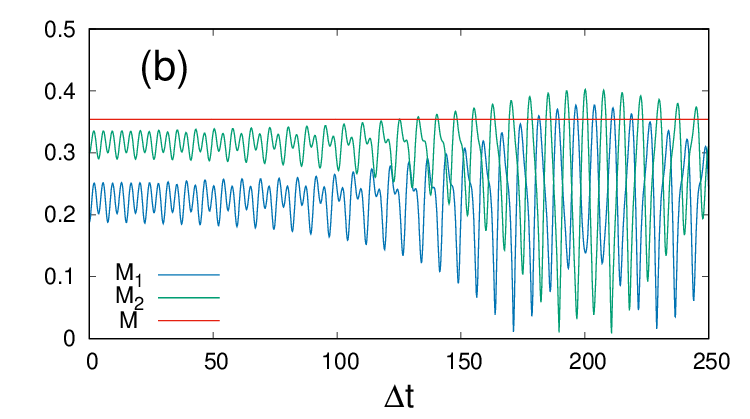, width=0.97\columnwidth}
\caption{
Example of dynamics with non-zero total angular momentum:
a) burst in function $I_1(t)$; b) absolute values
$M_1=|{\vec M}_1|$, $M_2=|{\vec M}_2|$, and $M=|{\vec M}|$.
}
\label{M1M2M}
\end{center}
\end{figure}

\section{Summary}

Thus, to describe the processes of radial tunneling of optical waves in
purely dielectric spherical Bragg resonators, we have derived the system
of ODE (\ref{C1_nonlin})-(\ref{C2_nonlin}). It demonstrates various dynamical
regimes, including nearly chaotic. The system is more simple than
the analogous system in the recent work \cite{R2026-2} where the tunneling
of light between two identical potential wells with spatially distant centers
was considered, within a model of effectively isotropic Epsilon-Near-Zero medium
in the presence of dissipation and external pumping \cite{R2026-1}. It was not
very difficult to consider in that model also the case of a centro-symmetric
potential $V(r)$ with two minima, derive the corresponding system of ODE for
radial tunneling, and then compare its predictions to the results of direct
three-dimensional (3D) numerical simulations. In such a way, a quite good
quantitative agreement has been obtained between the ODE system and the full
3D wave model. That result (not published yet) provides an indirect
evidence for correctness of equations (\ref{C1_nonlin})-(\ref{C2_nonlin}),
since a direct numerical simulation of the 3D equation (\ref{E_eq_nonlin})
seems to be a quite heavy task at present moment. Indeed, the required
numerical resolution should be as fine as $600^3$ points (about 30 wavelengths
along a spatial dimension, and about 20 points per one wave). Besides that,
in this case the computation should cover a long dimensionless time period
as $10/\Delta \sim 10^4$. 

The analysis can be generalized to the case of several ($N_s\geqslant 2$) shell
modes with $l=1$, which are concentrated on different radial coordinates.
In this case, as before, the nonlinearity appears in the equation for the vector
amplitude of the central mode only, but the linear part of the system is
expressed through an orthogonal ``localization matrix'' $\hat S$ of size
$(1+N_s)\times (1+N_s)$:
\begin{equation}
i\dot {\bf C}_\nu = [\hat S\,\mbox{Diag}\{\Delta_\kappa\}\,\hat S^{T}]_{\nu\mu}
{\bf C}_\mu +\delta_{\nu,1}[\partial H^{(4)}/\partial {\bf C}^*_1],
\end{equation}
where $\kappa,\mu,\nu =1,\dots,(1+N_s)$, and $\Delta_\kappa=(\xi_\kappa -\xi_1)$.

Finally we note that a doubtless advantage of a purely dielectric construction
is in its very small dissipation, so that no pumping may be needed to observe the
light tunneling occurring at typical time intervals like $10^4$ wave periods.

\vspace{1mm}

{\bf Funding.}
This work was supported by the Ministry of Science and Higher Education
of the Russian Federation, state contract no. FFWR-2024-0013.

\vspace{1mm}

{\bf Conflict of interest.}
The author of this work declares that he has no conflict of interest.

\end{document}